# The effect of liquid on the vibrational intensity of a wineglass at steady state resonance

Junghwan Lee

*CheongShim International Academy, Gyeonggi-do, Republic of Korea*

**Abstract** – As a liquid is inserted into a wineglass, the natural frequency of the wineglass decreases. This phenomenon, known as pitch lowering, is well explained in past papers. However, previous literature have not yet mentioned that pitch lowering also reduces the resonance intensity of a wineglass. Thus, this present paper aims to extend the body of research on this topic by describing the relationship between pitch lowering and its effect on resonation intensity. To do so, we identify the vibrating wineglass wall as a damped harmonic oscillator, derive a theoretical model, and find that the resonance intensity of the wineglass is proportional to the square of its natural frequency, under the assumption that damping stays constant. However, our experiments showed the coefficient of damping to increase with respect to the amount of liquid, which caused the data to deviate from its theoretical predictions. We conclude by discussing the accuracy and limitation of our proposed model.

## 1  Introduction

Musical wineglasses, also known as water xylophones, are excellent sources of entertainment for both children and adults. With various volumes of liquid inside wineglasses of the same geometry, one can manipulate the natural frequency of each wineglass to create an octave or more while maintaining aesthetical uniformity. Spectacular and straightforward, this mini instrument has been the subject of informal musical performances both on and off the dinner table.

But it's time to get into the physics of this musical wineglass. How exactly does liquid play in affecting the sound produced by a vibrating wineglass? An empty wineglass pushes air while it vibrates, but a wineglass partially filled with liquid pushes a mix of air and liquid. Since the mix of air and liquid is heavier than plain air, the partially filled wineglass uses more energy per vibration, and as a result, the natural frequency of the system, as well as its vibrational intensity, decreases because it vibrates less and weakly.

As interesting of a phenomenon it is, preceding researchers [1–5] have looked upon the effect of the liquid on the natural frequency of the wineglass. They labeled this phenomenon as 'pitch lowering' because the presence of liquid proportionally lowered the resonant frequency of the wineglass [1]. Using a simplified cylinder model for a wineglass, French formulated an equation that models the fall-off of the natural frequency with respect to the height and properties of the wineglass and the existent liquid [2]. Courtouis [1] and Chen [3] then further developed the model for pitch lowering by stating that the spatial distribution liquid inside the wineglass plays a crucial factor in determining the new resonant frequency, as opposed to volume.

Researches on the modes of a resonating wineglass have been conducted as well. Skeldon identified the vibrational structure of a wineglass as two orthogonal quadrupole modes and found that increasing harmonics result in the wineglass top to undergo star-shaped oscillation with growing points, but with a decrease in amplitude [4]. Extending this finding with liquids, Junct showed that the presence of liquid does not affect the vibrational structure of the wineglass [5].

However, not only does the existence of liquid lower the natural frequency of the wineglass, but, as observed in this paper, it also contributes to the reduction of the wineglass's resonance intensity.

To find a relationship between the pitch lowering phenomenon and the resonance intensity of a wineglass, we present a solution through theoretical and experimental work. The solution presented is for a wineglass at steady state resonance.

## 2  Observation

The process of resonating a wineglass consists of an external sound wave propagating in the direction of the wineglass. When the frequency of the sound wave is of similar value as the wineglass's natural one, the wineglass becomes a driven oscillator due to the sound wave. It vibrates in resonance and in the process, gives off its own sound wave.

In order to resonate a wineglass under controlled environments, we set up an experiment as shown in Figure 1. The setup includes a loudspeaker sourcing sinusoidal sound waves at the direction of the wineglass and a sound level meter detecting sound waves by its amplitude and frequency. Specific model names for the equipment used in this experiment is as follows: loudspeaker Tannoy Stirling SE, free-field microphone B&K type 4191, and audio analyzer APx525 by Audio Precision. The free-field microphone measures the sound pressure in Pascal, but with the help of the audio analyzer, the Pascal units are converted into the unit of sound power (Watts) by converting the measured pressure (Pascal) into dBSPL units then into dBm units, which can be expressed into Watts.

The wineglass used was of cabernet geometry with a bowl of 10 cm height and 6 cm rim diameter. Experiments were conducted in an anechoic chamber.

Firstly, we found the natural frequency of the wineglass by printing a power spectral density while rubbing the wineglass's rim with a wet finger and analyzing the highest peak. Then, in order to observe the effect of harmonics on the vibrational intensity of a wineglass, we sent out a sinusoidal wave of identical frequency as the wineglass's natural one through a loudspeaker. The results, in the form of a power spectral density, are as shown in Figure 2.

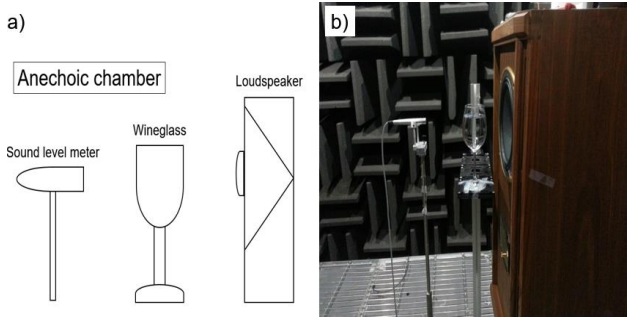

**Figure 1:** Schematic (a) and photo (b) of the experimental setup

As seen in Figure 2, we found out that while this wineglass resonates, the amplitude of its harmonics are approximately two-thousandth of the amplitude at its fundamental frequency. Because of such large differences, the effect of harmonics on the resonance structure of the wineglass is seemably negligible. The mechanics of a resonating wineglass, at its fundamental mode, then takes the shape of a quadrupole mode [6] vibration as shown in Figure 3.

## 3    Theory

The movement of its walls characterizes the vibration of a resonating wineglass. A point on the wall, when the wineglass is in perfect resonance due to an external force, is a damped harmonic oscillator. A mathematical notation is as follows.

$$m\ddot{x} = -kx - b\dot{x} + F_L \sin(\omega t) \tag{1}$$

Where $m$ is the effective mass, the mass related to the oscillation; $x$ the displacement of a point on the wineglass wall; $k$ the elasticity coefficient; $b$ the damping coefficient; $F_L$ the modal projection of the incident pressure field onto the fundamental eigenmode produced by the loudspeaker; $\omega$ the frequency of the external force; and $t$ the time. $x$ is mathematically defined as $x = A\sin(\omega t)$, where $A$ is the amplitude of displacement as depicted in Figure 3.

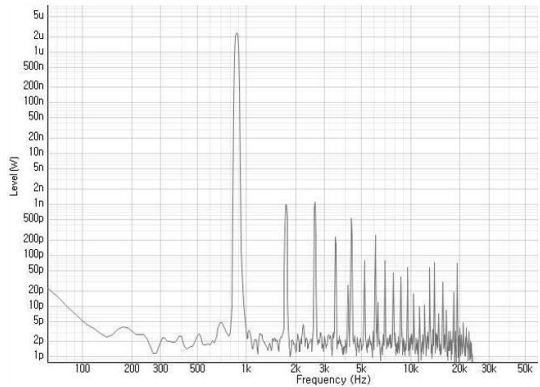

**Figure 2:** A power spectral density of the sound produced by the wineglass as it was forced to resonate. The peaks show the amplitude of the fundamental frequency, as well as its harmonics.

In addition, because our experimental procedure (rubbing the rim with a wet finger) gives us knowledge of the wineglass's natural frequency, the loudspeaker was programmed to send out a sound wave of frequency equal to the natural frequency of the wineglass in order to create perfect resonance. This leads us to set $\omega_0 = \omega$.

When a wineglass in steady state resonance, the damping force and driving force from Equation 1 equates each other. Equation 2 express this characteristic.

$$b\dot{x} = F_L \sin(\omega t) \tag{2}$$

With these particular characteristics identified, we then go on to find the vibrational intensity produced by the wineglass at steady state resonance. To do so, we start by modeling the power absorbed by the wineglass.

The power absorbed by the wineglass, $P_A$, by the external force is modeled as follows.

$$P_A = F_L \sin(\omega t) \cdot \dot{x} \tag{3}$$

Using Equation 2 and substituting $F_L \sin(\omega t)$ with $b\dot{x}$ gives

$$P_A = b\dot{x}^2 \tag{4}$$

Substituting for $\dot{x}$, we get

$$P_A = b(A\omega_0 \cos(\omega_0 t))^2 \tag{5}$$

Since $P_A$ is a function of time, we find the maximum value of the power absorbed by the wineglass: $\overline{P_A}$.

$$\overline{P_A} = bA^2 \omega_0{}^2 \tag{6}$$

The average power and the amplitude of power are equal to half the maximum value.

Now that we know $\overline{P_A}$, we need to find its relationship with the power of the soundwave from the loudspeaker and the power exerted as sound from the wineglass. Intuitively, the values of the three powers must be directly proportional to each other, but to verify, we took experimental methods. Using the same setup in Figure 1, three different sound waves of power (1 mW, 5 mW, and 10 mW) were produced from the loudspeaker at nearby frequencies to the wineglass's natural one (865 Hz).

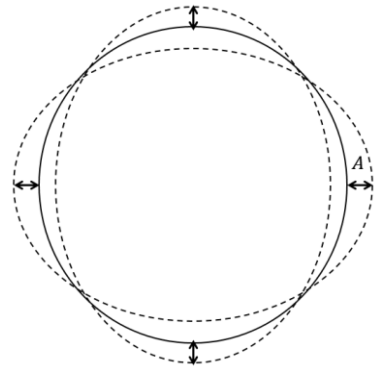

**Figure 3:** Sketch of a quadruple mode vibration on a wineglass rim, as seen from above. The solid line represents a still rim while dotted lines represent the vibration of the rim.

$\overline{P_L}$ indicates the maximum power of the sound produced from the loudspeaker and $\overline{P_W}$ the maximum power of the sound produced from the wineglass. $\overline{P_L}$ was measured by removing the wineglass in the setup illustrated in Figure 1 and $\overline{P_W}$ was measured by subtracting $\overline{P_L}$ from the total acoustic field detected by the free-field microphone.

Results are as shown in Table 1.

**Table 1:** The maximum power of the soundwave exerted by a resonating wineglass ($\overline{P_W}$) due to external sound waves of varying power and frequencies. Units are in milliwatts (mW).

| Frequency of sound produced from the loudspeaker (Hz) | $\overline{P_L}$ (mW) | | |
|---|---|---|---|
| | 1 | 5 | 10 |
| 830 | 0.12 | 0.62 | 1.28 |
| 840 | 0.32 | 1.62 | 3.30 |
| 850 | 0.42 | 2.06 | 4.28 |
| 860 | 0.46 | 2.24 | 4.68 |
| 865 | 0.46 | 2.34 | 4.72 |
| 870 | 0.42 | 2.14 | 4.46 |
| 880 | 0.20 | 0.96 | 1.98 |
| 890 | 0.02 | 0.12 | 0.24 |
| 900 | 0.00 | 0.00 | 0.00 |

Table 1 shows that the maximum power of both the sound produced from the loudspeaker and the wineglass are directly proportional to each other. We reasonably infer that the maximum power absorbed by the wineglass, which is an intermediate, follow the trend of direct proportionality. We then model this relationship as

$$\overline{P_W} \propto \overline{P_A} \propto \overline{P_L} \qquad (7)$$

Under the assumption that the coefficient of damping remains independent to liquid parameters, we substitute $\overline{P_A}$ with its relationship to $\omega_0$ in Equation 6 into Equation 7 to obtain an equation with respect to $\overline{P_W}$.

$$\overline{P_W} = RbA^2\omega_0{}^2 \qquad (8)$$

Where $R$ is the linear proportionality ratio between $\overline{P_W}$ and $\overline{P_A}$. Equation 8 states that the power exerted by the wineglass is proportional to the square of its natural frequency.

Table 1 shows that the maximum power of both the sound produced from the loudspeaker and the wineglass are directly proportional to each other. We reasonably infer that the maximum power absorbed by the wineglass, which is an intermediate, follow the trend of direct proportionality. We then model this relationship as

$$\overline{P_W} \propto \overline{P_A} \propto \overline{P_L} \qquad (7)$$

Under the assumption that the coefficient of damping remains independent to liquid parameters, substituting $\overline{P_A}$ with its relationship to $\omega_0$ in Equation 6 into Equation 7 gives

$$\overline{P_W} \propto \omega_0{}^2 \qquad (8)$$

Which states that the power exerted by the wineglass is proportional to the square of its natural frequency.

## 4   Results and Analysis

To test the theoretical model proposed in Equation 8, we needed to obtain the values of the maximum power exerted by the wineglass with respect to its natural frequency. To find the natural frequencies of wineglasses with various liquid parameters, we printed a power spectral density while rubbing the wineglass's rim with a wet finger and took the frequency of the highest peak. Three different types of liquids (oil, water, and honey) were inserted in the same wineglass to four different heights (2.5 cm, 5 cm, 7.5 cm and 10 cm) to create a total of 12 different combinations. The density of oil was 0.92 g/mL, water 1.02 g/mL and honey 1.50 g/mL. Results are as shown in Table 2 and Figure 4.

**Table 2:** Natural frequencies of a wineglass filled with various types and quantities of liquid. Units are in hertz (Hz).

| Type of liquid | Height of liquid (cm) | | | | |
|---|---|---|---|---|---|
| | 0 | 2.5 | 5 | 7.5 | 10 |
| Oil | 865 | 865 | 798 | 632 | 457 |
| Water | 865 | 855 | 797 | 610 | 445 |
| Honey | 865 | 860 | 760 | 540 | 385 |

As Figure 4 shows, the relationships between the height of the liquid and natural frequency are non-linear. Since the circumference of the wineglass increases nonlinearly with respect to its height, this result agrees with findings from preceding papers which stated that spatial distribution plays a more prominent role in pitch lowering as opposed to the height and volume of the liquid [1, 3].

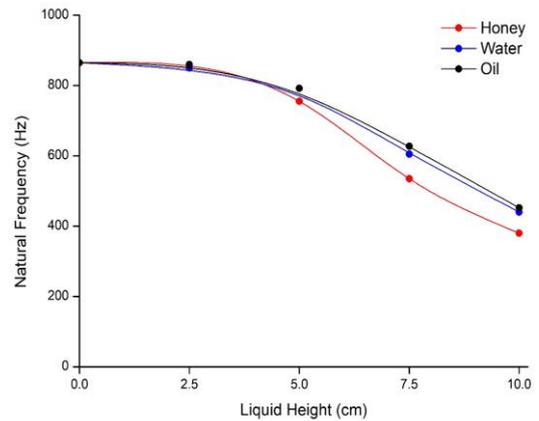

**Figure 4:** Graphical representation of data from Table 2.

We now turn our focus to the relationship between the natural frequency and the resonance intensity of the wineglass. Using the experimentally found natural frequencies, we recorded the maximum power exerted by the wineglass for every variation when the loudspeaker exerted sounds with a 10 mW power at frequencies similar to the wineglass's natural one. Figure 5 shows the maximum power exerted by each of the variant wineglasses to wave frequencies similar to its natural one.

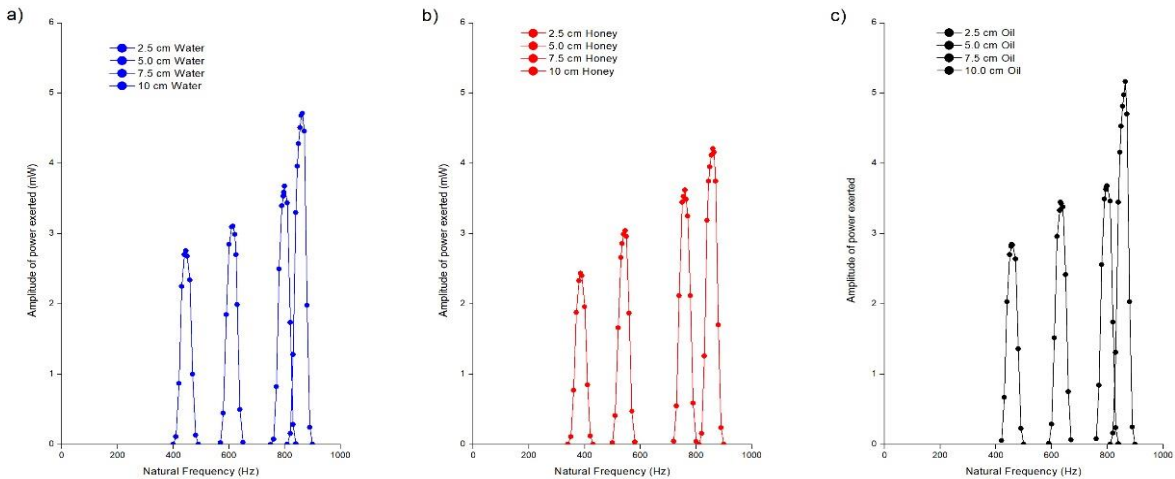

**Figure 5:** Resonance curves for each wineglass variant. The maximum power exerted by the wineglass was tested at near resonant frequencies and connected using a straight line. (a) is the resonance curve for wineglasses filled with water, (b) oil, and (c) honey. The data points for each liquid height, from left to right, are 10 cm, 7.5 cm, 5 cm, 2.5 cm for all three graphs.

Taking the peak values of each resonance curve in Figure 5, which indicate points of perfect resonance, we plotted the data on a scatterplot and posed a quadratic fit with only a second term ($y = ax^2$ fit) for each wineglass variance to test Equation 8. Results are as shown in Figure 6.

We notice that the quadratic fit for the data on three different types of liquids are approximately equal to each other in Figure 6. Moreover, the combination of data from all three liquids follow a trend altogether. We can thereby conclude that the different types of liquids and the differences they have, such as viscosity and density, are all factored in through pitch lowering.

For wineglasses with natural frequencies similar to 800 Hz (0~5 cm liquid height), the data points fit closely with the regression curve while data for wineglasses with lower natural frequencies (7.5~10 cm liquid height) deviate upwards. This is because we assumed that the coefficient of damping stays independent from liquid when we modeled Equation 8. The wineglasses with lower frequencies are accordingly the wineglasses with a higher liquid height, and as the liquid gets in contact with more surface area of the wineglass, the extra mass of the liquid and surface tension act to increase the coefficient of damping. Therefore, parallel with what is expected from Equations 6 and 7, the maximum power exerted from the wineglass attains a higher value than what the fitted curve suggests.

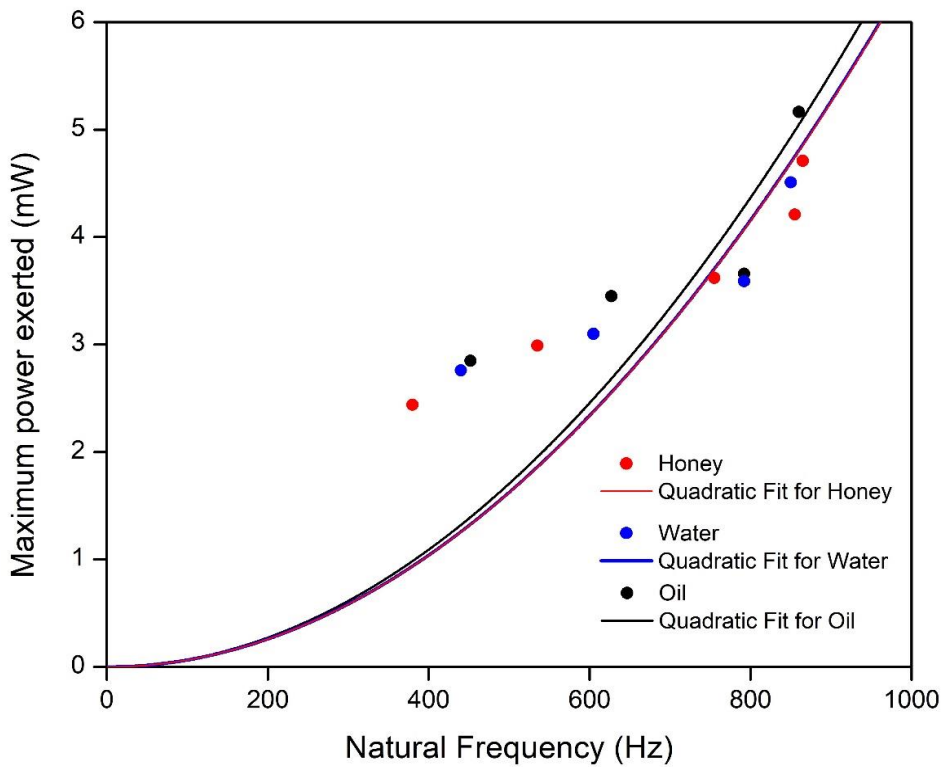

**Figure 6:** The maximum power exerted by the wineglass, with different variances in liquid, as it resonates. The quadratic fit for honey and water overlap each other. The leftmost three data points correspond to a wineglass with honey, water and oil filled up to 10 cm. The next three points correspond to 7.5 cm, 5 cm and so forth. The rightmost point is when no liquid was inside.

# 5 Conclusion

Through experimental work, we found results that complement previous findings, and extended the research by connecting it to the resonance intensity of a wineglass at steady state resonance. Regarding liquids, pitch lowering, and the resonance intensity of the wineglass, parameters that vary between different types of liquid, such as viscosity and density, do not have additional effects on the system after it has contributed to pitch lowering. With that in mind, we took the pitch lowering phenomenon as given and investigated its effect on the resonance intensity of the wineglass. We found that for wineglasses with small volumes of liquid, the coefficient of damping takes on a near constant value, and thus a clear squared relationship exists between the natural frequency and the maximum power exerted by the wineglass. However, for wineglasses with larger volumes of liquid, the quadratic fit underestimates the maximum power exerted by the wineglass because the damping coefficient starts to increase due to liquid. Nevertheless, the proportional relationship between the maximum power exerted by the wineglass with respect to damping and natural frequency in Equation 6 stands correct, though Equation 8 can only be used for wineglasses with small amounts of liquid.

Mind that the sound exerted by the wineglass is due to the vibration of the wineglass, and these findings apply directly to its vibrational intensity. Moreover, the power exerted by the wineglass as sound is directly proportional to its sound intesity, which uses the familiar units of decibels.

Future work could use a high-speed video camera to determine the amplitude of displacement and compute the coefficient of damping. With the value of the coefficient of damping, the resonance intensity of a wineglass at transient stages could be modeled, and even possibly a model between resonance intensity and various liquid parameters.

## Acknowledgements

I would like to thank my father Mr. Sangup Lee who has helped me gain access to the anechoic chamber for my experiment, physics teacher Mr. Jaesung Yoon for discussing ideas together, and anonymous reviewers for comments that have helped to improve my paper.